\documentclass[%
 reprint,
nofootinbib,
 amsmath,amssymb,
 aps,
floatfix,
]{revtex4-1}

\usepackage{graphicx}% Include figure files
\usepackage{dcolumn}% Align table columns on decimal point
\usepackage{bm}% bold math
\usepackage{appendix}
\usepackage{hyperref}
\usepackage{adjustbox}
\usepackage{xcolor}

\newcommand{\ba}{\begin{eqnarray}}
\newcommand{\ea}{\end{eqnarray}}

\newcommand{\be}{\begin{equation}}
\newcommand{\ee}{\end{equation}}

\begin{document}

%\preprint{APS/123-QED}
%\preprint{LTH-1291}

\title{The Quest for Understanding\\ 
The Case of the Upgraded Superconducting Super Collider}

\author{Alon E. Faraggi$^{(a,1)}$}
%\altaffiliation[Also at ]{Physics Department, XYZ University.}%Lines break automatically or can be forced with \\
%\author{Second Author}%
\email{$^{1}$Alon.Faraggi@liverpool.ac.uk}

\affiliation{$^{(a)}$Department of Mathematical Sciences University of Liverpool, Liverpool L69 7ZL, UK. }

\date{\today}% It is always \today, today,
             %  but any date may be explicitly specified

\begin{abstract}
\begin{description}
\item[Abstract]
Fundamental particle physics is at a cross road. On the one hand
the Standard Model successfully accounts for all experimental observations to date. On the other hand the ElectroWeak symmetry 
breaking mechanism is poorly understood and suggests the 
existence of new physics within reach of future colliders. 
Building on LHC experience, a hadron collider using the well 
established LHC magnet technology in a 90--100km circular ring,
can reach the required 50--60TeV energy range and uncover
the next layers of reality by the early 2040s.
Building on CERN experience, it is envisioned that such a facility 
could be established as a Middle East collaboration at the SESAME site
and funded as a regional and international project.

%\item[Pacs numbers] 12.38.-t,12.38.Bx,12.38.Aw
\end{description}
\end{abstract}

% PACS, the Physics and Astronomy
                             % Classification Scheme.
%\keywords{Suggested keywords}%Use showkeys class option if keyword
                              %display desired
\maketitle

%\tableofcontents

%\section{Introduction}\label{sec:Introduction}

Since early times human beings have had the need to explain how the world around them came to 
be and the rules that govern it. The biblical story of creation tells us that the world 
was created in six days. Make what you will of the biblical story, the fact is
that it testifies to that need, and documentation of other early human 
societies give us further evidence to that basic need. 
Since the early sixteen century there was a twist in the tale. 
Up to that point the world to explain existed, and its explanations 
were primarily qualitative. 
%, though quantitative modelling existed as well. 
But from that point, human beings increased the development of 
experimental methods and the use of mathematics to describe the outcome of
these experiments. This cumulated into what we now call the scientific 
method. 

Another theme on the quest for understanding the world around us is that of unification. 
Physics is first and foremost an experimental science. But the language 
that we use to describe the experimental data is mathematics. 
It therefore makes sense to build mathematical models that describe 
wider and wider range of experimental observations. The successful mathematical models
are those that are able to account for the widest range of experimental data.
A comment is in order here. 
One often hears on popular blogs, and sometimes also among esteemed 
colleagues, that a physical theory should make predictions. A physical 
theory should be predictive. A more appropriate characterisation is to say that 
a physical theory should be calculable. In a physical theory, 
given some initial conditions, we should be able to calculate some outcomes
that can then be observed or not observed in some experiments. We can then 
say that the physical theory is testable. We should, however, express some 
caveats here. The first is the research process that goes into formulating
physical theories. In that process we may consider possible theories 
that are in fact not predictive in the sense described above. Ultimately, 
those theories will not be acceptable physical theories. The second is that 
some predictions of theories that may eventually prove to be good physical 
theories, may give rise to predictions that may not be tested at the time of 
their formulation. Their acceptance as good physical theories may have to await the
development of new experimental instruments and methodology. A good historical example
is that of the heliocentric versus the geocentric model of the solar system. A heliocentric
model of the solar system
was proposed by Aristarchos of Samos in the third century { BC} versus the geocentric 
model that was championed by Plato, Aristotle and Ptolemy. Yet, not until the development 
of the telescope by Galileo that decisive observations were made
that could settle the question. 
Galileo is often credited as the father of the modern scientific method in the sense alluded 
to above, of using mathematical modelling to describe physical phenomena and subjecting
the theoretical predictions of the mathematical models to observational tests. 
In contemporary science a mathematical model cannot be accepted as a physical theory
unless its predictions can be subjected to such experimental verification. 

The theme of unification is a prevailing theme through the ages. Newton showed that
the celestial motion of the planets in the solar system and the terrestrial motion of rigid bodies 
on earth are governed by the same mathematical laws. Maxwell unified the electric and magnetic
forces into one set of equations. Closer to our times, Dirac unified
special relativity with quantum mechanics; in the 1960s
Glashow, Salam and Weinberg 
unified the Quantum Electro Dynamic (QED) theory that describes the 
relativistic electromagnetic interaction which is long range, with the
subatomic weak interaction which is short range. 
This cumulated into the Standard Model of the Electro Weak interactions. 
During the 1970s, Gross and others developed the theory of
Quantum Chromo Dynamics (QCD) that describes the strong interaction
that binds the nuclei together, and was augmented to the Standard Model, 
to describe the three forces that operate in the subatomic 
world: the strong; the weak; and the electromagnetic. 
All three forces are formulated mathematically by the gauge 
principle in which the interaction between particles is described
as the exchange of another particle, which is the mediator of the 
interaction. In parallel to the development of the gauge 
theories of the interactions, the zoo of elementary particles 
was discovered. The process of discovery started 
in 1897 with the discovery of the electron, the first elementary particle 
to be discovered, by J.J. Thomson in the Cavendish laboratory
in the United Kingdom. In the next 100 years the process of 
discovery continued unabated. This is not the place to give a proper
account of this remarkable development and achievement in the 
history of human ingenuity. A good reference that describes some of
the aspects of this story is the book ``The hunting of the Quark", by 
Michael Riordan \cite{Riordan}.
The last elementary matter particle to be discovered was
the top quark that was discovered in the Fermi laboratory in the United States.
The top quark was the last elementary matter particle to be discovered, 
but it was not the last particle to be discovered. That honour 
belongs to the Higgs particle that was discovered in 2012 in the 
CERN laboratory in Switzerland. The Higgs particle has 
the privilege of enabling the marying of the electromagnetic interaction,
which is long range and mediated by a massless force mediator, and the
weak interaction, which is short range and mediated by massive 
force mediators, in one mathematical framework. The mathematical 
formulation of the Standard Model of the subatomic elementary particle and 
interactions crystallised by the mid-1970s. The experimental 
affirmation of this mathematical parametrisation of the observational data
is a process that continues in current and planned 
experiments. The Standard Model 
is an effective mathematical theory, albeit an extremely successful 
one, and is bound to cease to provide a viable effective theory
at some energy scale above the scale which is accessible to contemporary
experiments. 

There aren't enough words of praise and admiration to do justice to
the achievement in the experimental discovery of the various bits
and pieces that make up the Standard Model and the continuing experimental
work that aim to test and explore the validity of the Standard Model 
and its possible theoretical extensions. The diamond in the crown 
is the Centre European Recherche Nuclear (CERN) near Geneva 
in Switzerland. It is testimony to the remarkable 
technical and social achievement of the work of thousands of 
scientists and engineers from all around the globe. 
It reflects on the ability of human beings to put aside their 
different cultural backgrounds and work toward a common goal. 
On the technical side the current Large Hadron Collider 
accelerator complex consists of 27km of superconducting 
magnets all tuned to work in tandem to keep the accelerated
protons to collide at specified points inside the
detectors. Each detector experiment is a marvel in itself
and combines the efforts of collaborations with  $\sim3000$ 
members. The technical ingredients that go into the detectors
are the Ferraris of the future. It is at the forefront of
technological achievements and developments. Nearly 4500 
years ago the great pyramids in Egypt were build and were 
the technological pinnacles of their time. The collider
experiments at CERN and at other elementary particle 
laboratories are the pyramids of our time. I invite every
reader to visit the laboratory and its new outreach complex, 
which is fantastical in itself. Since 1996 I have been a 
frequent visitor to CERN, visiting every Summer for typically
periods of two weeks, though I have also stayed for longer. 
I wish it and its dedicated personnel to continue to 
lead the field of experimental particle physics. 
It is a marvel to behold. 

The establishment of the Standard Model as 
providing the correct parametrisation of all subatomic 
experimental observations opened the door to the 
unification of all the three subatomic forces into
one fundamental force at an energy scale far removed 
from the Standard Model energy scale. These are known as
Grand Unified Theories (GUTs) and were mostly developed from
the mid--1970s until the mid 1980s, though research on their
mathematical structure and implications continues to this 
day. There is, however, one other force which is not included 
in GUTs. That of gravity, which was the first force to
be recognised and formulated mathematically. However, 
as a fundamental force between elementary particles 
there is a problem. The problem of infinities. For a
physical theory to be useful as a calculational or
predictive framework it should be free of infinities. 
Otherwise, their predictions cannot be trusted. 
Infinities also arise in the mathematical formalism 
of the subatomic interactions. However, the 
number of infinities is small. What we do then is absorb
the infinities into a number of parameters that we 
measure experimentally. This is essentially sweeping 
the problem under the rug. We cannot calculate the 
physical parameters from some mathematical principle, 
but we can measure and determine them experimentally. 
We can then use these measured parameters to calculate numerous
other experimental observables. We say that the theory is 
renormalisable and predictive. The problem with gravity 
is that there is an infinite number of infinities.
We then say that the theory is not renormalisable and 
not predictive. It is therefore not possible to write 
a fully consistent theory of gravity which is unified
with the Standard Model of particle physics in a theory in
which the elementary particles are represented as 
elementary point particles. The problem lies in the fact that
gravity is a theory of spacetime itself and idealising particles 
as point particles results in an infinite number of infinities and
the formalism breaks down. Another theory that has been 
developed in earnest since the mid--1980s, called string theory,
comes to the rescue. In string theory elementary particles are 
described as strings rather as idealised point particles. 
The result is that the theory is finite to begin with and
no infinities arise. In string theory it is indeed possible to 
find solutions that mimic the Standard Model of particle 
physics as well as the equations that govern the gravitational 
interaction, yet the theory is finite and predictive. However, 
a warning is in order here. The string unification occurs at
a scale which is far removed from the Standard Model scale 
and there is no way at present to probe the string hypothesis
directly. 

\begin{figure} 
\begin{center}
\includegraphics[width=7.cm]{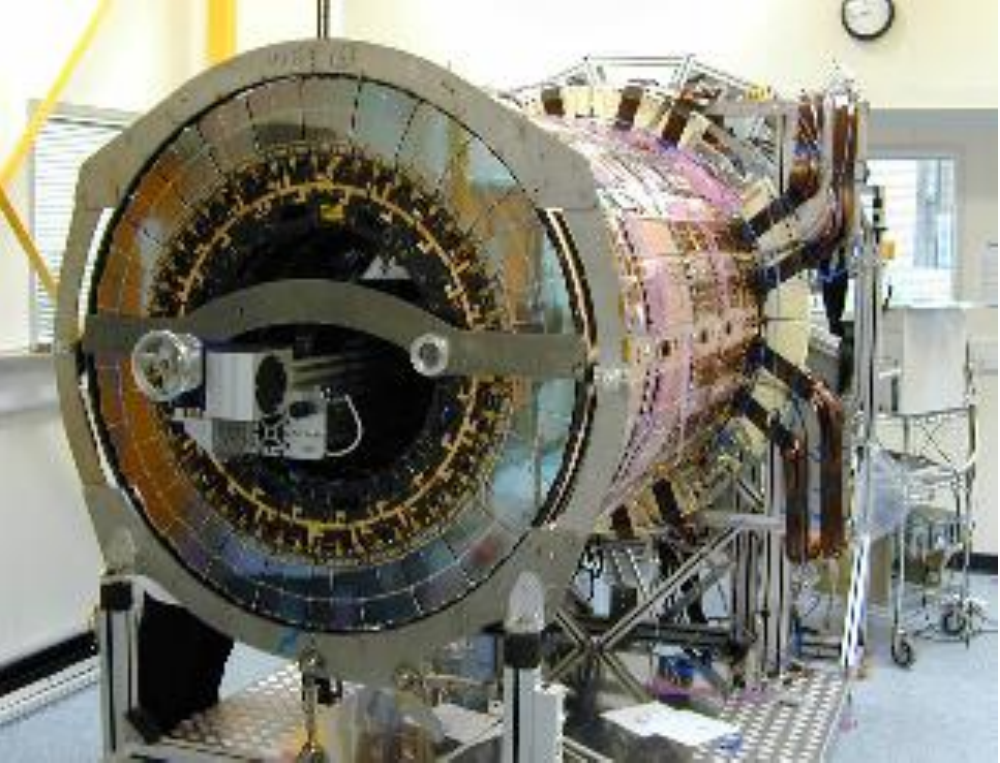}
\caption{Forward silicon central tracker of the ATLAS experiment at the LHC. }
\label{endcap}
\end{center}
\end{figure}

\begin{figure} 
\begin{center}
\includegraphics[width=7.cm]{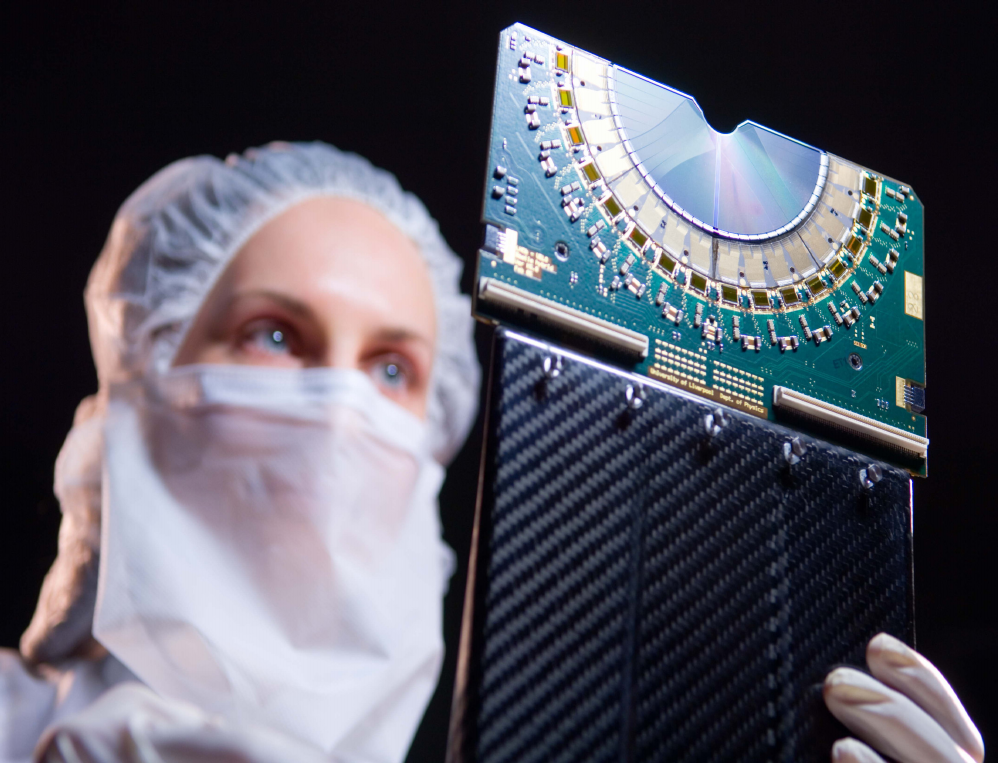}
\caption{Vertex detector of the LHCb experiment.}
\label{lhcbvelo}
\end{center}
\end{figure}

It would seem that we are near the end of the road to 
describe all the fundamental matter and interactions in
a single mathematical framework. However, this could not 
be more misleading. Our mathematical description of
the fundamental matter and interactions faces 
some very severe problems. The first and perhaps the 
most perplexing is called the hierarchy problem. The mass 
scale of the Higgs particle is of the order of
$10^{-25}$Kg. However, the GUT mass scale where all the
subatomic interactions can be unified is of order 
$10^{-11}$Kg, {\it i.e.} fourteen orders of magnitude 
separate the two scales. For comparison we can compare the 
size of human beings, which is roughly $1$ meter to that of the 
atom, which is roughly $10^{-10}$ meters, {\it i.e.}
ten orders of magnitude. Thus, the separation between the 
Higgs mass scale and the GUT mass scale is four orders 
of magnitude bigger than the separation between the size
of human beings and the size of the hydrogen atom. A vast 
separation indeed. The problem is that the Higgs mass scale 
should not be separated at all from the GUT scale. For the 
other elementary particles, the matter particles and the 
force mediator particles, their mass scales are protected 
by symmetries. But the Higgs mass scale is not protected
by a symmetry. This puzzle is known as the hierarchy problems
and is perhaps the most perplexing problem in our
understanding of the physical world. In truth, we do not really 
understand the nature of the Higgs particle and the 
mechanism that allows for the massless and massive force 
mediators to exist together in the Standard Model. 
Several solutions have been proposed to this problem 
that posits the existence of a new symmetry called supersymmetry,
or that the Higgs particle is not an elementary particle 
after all, but is itself composed of more elementary particles
that are bound by some new strongly interacting force. In reality,
what we need is new experimental input to shed light on the 
mass scale above the Higgs mass scale. 
There are numerous other problems in our current 
understanding of the physical world that we inhabit. 
Some have to do with the structure of the Standard Model
and others have to do with the synthesis of the Standard 
Model with gravity, {\it e.g.} the dark matter and dark 
energy problems. But the hierarchy problem is the one 
that calls for the contruction of a new collider experiment
and for which a new collider experiment is the most appropriate
experimental tool. 
The Large Hadron Collider (LHC) at CERN is the current 
experimental instrument to probe the physics of the Higgs particle. 
The LHC has been a tremendous success that is further testimony to
human ingenuity. Not only did it discover the Higgs particle and 
elucidated its properties, but it continues to provide immeasurable 
quantities of new data in a highly complex experimental
environment. In particle physics length and mass
scales are typically referenced in energy units.
This is due to Einstein's famous formula that relates
mass and energy, $E=Mc^2$, where $E$ 
is the energy; $M$ is the mass; and $c$ is the speed of light. 
Additionally, for relativistic energies $E=|p|c$, where $p$ 
is the momentum and by Heisenberg uncertainty principle
$\Delta x\Delta p\ge \hbar$, meaning that larger momentum uncertainties
are associated with smaller distance resolutions. Hence, increasing
the energy scale of the collider experiment probes shorter and shorter distances 
in length scale. In fundamental physics all scales are measured in energy units and typically in electron Volts and their powers. 
The nuclear energy scale, the scale of the inner nucleus
inside the atoms, is of the order of 1GeV or $\sim10^{-15}$ meters. 
The LHC runs at $14\times 10^{12}{\rm eV} =14{\rm TeV}$
and will continue to run at this energy for the next 10--15 years
accumulating more and more data on the Standard Model processes 
and beyond. The experimental particle
physics group at the University of Liverpool 
participates in two of the LHC experiments. 
In collaboration with other institutions, 
it contributed one of the Silicon Central Trackers to the
ATLAS experiment, which is shown in figure \ref{endcap},
and the Vertex Detector to the LHCb experiment, a segment of which
is shown in figure \ref{lhcbvelo}.
If you can get one of the experimentalists in the physics 
department to sneak you in, you can see from outside the 
clean rooms where these pieces of kit were assembled. 
And then imagine, or see online, the images of collisions 
that these instruments aim to see and produce. Unraveling 
the inner cores of reality. Probing the realms beyond the
present level of human understanding. Seeking to understand
what lies beyond. A never ending quest for understanding.  

The LHC experiment at CERN will continue to provide gold
class data on the Standard Model processes and measure
its parameters to better and better precision. However, 
it will be limited in unravelling the next layers of 
reality. For that new experimental facilities are required. 
Furthermore, without further experimental input in the 
energy regime beyond the LHC scale real progress 
in understanding the mathematical formalism that
underlies the Higgs mechanism will not be possible. 
We can continue to entertain for eternity 
mathematical models beyond the Standard Model, 
but to know whether any of them is relevant in the 
real world will not be possible without new experimental
input. 
The question of the experimental collider facilities beyond the
LHC has been intensely examined by the three main international
players in this area over the past few years 
\cite{Butler:2023eah, Gourlay:2022odf, EuropeanStrategyGroup:2020pow}. 
The efforts are international in spirit. Particle physics
experiments are international enterprises with CERN leading 
the fray, setting the example, and establishing the 
practices and procedures. 
The efforts can be divided into the regions that will 
host specific facilities. The three main theatres are 
China, Europe and the United States. The future
facilities in the energy and precision frontier
can be divided into hadronic and leptonic machines, 
where in hadronic machines protons are accelerated, 
whereas in the leptonic machines electrons and 
positrons are accelerated. The leptonic
machines offer better precision. The electrons
and positrons are elementary point particles, whereas
the protons are composed of quarks that are 
confined by the strong interactions inside the proton.
Our knowledge of the energy and momentum of the
colliding electrons and positrons is precise whereas
that of the colliding quarks inside the proton is murky. 
The tradeoff is that the electron is much lighter 
than the proton and we can accelerate protons to 
much higher energies. Leptonic colliders will therefore
be able to measure the properties of the Standard Model
Higgs sector more precisely than hadronic colliders, 
whereas hadronic colliders can reach higher energies 
and have better prospects of uncovering new physics. 

The other major consideration in the planning of 
the future machines is the type of magnets 
that will be used, in particular in hadronic 
accelerators. The LHC tunnel currently in operation in Geneva
is a 27km circular tunnel. The main magnets are
made of superconducting Niobium--Titanium (NbTi) alloys. 
They operate at temperature of 1.9K and produce a 
magnetic field of 8.3 Tesla. The design and 
manufacturing of the magnets is one of the key elements
in both the success and the cost of a collider experiment. 
Future colliders in planning are projected to operate magnets
with stronger magentic fields, of the order of 16 Tesla. 
The current alloy technology cannot sustain stable 
magnetic fields of this strength so new alloys need to be developed. The plan is to construct 
superconducting magnets that are based on the use
of Niobium--Tin (${\rm Nb}_3{\rm Sn}$) alloys 
to reach operating fields of 16T. However, as of today, 
the technology has not yet been fully developed. Like any 
such technical design, we cannot be fully confident 
that the magnets will work or it may take longer 
than expected to produce stable magnets that can
be manufactured on industrial scale and sustain a stable 
magnetic field over the length of the accelerator and 
its envisioned operating period. 

In Europe the CERN High--Luminosity LHC will operate until 
the late 2030s at 14TeV Centre of Mass (CoM). During that 
period a 90--100km circular tunnel will be dug for the 
Future Circular Collider (FCC). The 
initial plan is to build an electron--positron collider 
FCC--$e^+e^-$ that will operate initially at 250GeV, twice
the Higgs mass, and at later stages at 350GeV and 500GeV. 
The FCC--$e^+e^-$ will be operational from the early to mid--2040s 
and will provide precise measurements of the
Higgs parameters. The FCC will then be converted to an hadronic 
FCC--$hh$ collider at 90-100TeV that will be operational from the 
early 2070s. The FCC--$hh$ will be a discovery machine that
will be optimal for probing the physics Beyond the Standard Model
(BSM). 

The US conducted an in depth exercise to examine its future
particle physics program that cumulated in the Snowmass 2021
report \cite{Butler:2023eah}. The immediate US collider 
physics program will continue its current focus on neutrino
physics and the long term recommendation of the 
Particle Physics Project Prioritization Panel (P5)
is to build a muon collider. The muon has identical properties
to the electron but is about 200 times heavier. The 
advantage is that a smaller ring can be used to reach higher
energies. The caveat is that the muon is not stable and 
has a lifetime of about $2.2\times 10^{-6}s$. This poses 
a major technological challenge. It would be fair to 
say that this is an open ended project which may become
operational as a physics machine in 50--100 years. 

The Chinese plans mirrors that of the Europeans with an
envisioned 90--100km tunnel that will house initially
a $e^+e^-$ lepton collider, the Chinese Electron--Positron Collider
(CEPC) with a second subsequent phase in the same 
tunnel as a hadron collider at 90--100TeV, 
the Super Proton--Protton Collider (SPPC). 
The physics case is similar to the FCC one
with the CEPC performing precision measurements 
of the Higgs parameters, whereas the SPPC will
be a discovery machine to probe the physics 
that lies behind the Higgs mechanism. The envisioned timeline
of the CEPC is within 10-15 years {\it i.e.}
it may become operational in the 2035--2040 period, 
with the SPPC becoming operational in the early 2060s. 
The word of caution of order here is that the technical
know how truly only exist at present at CERN and any new 
collider experiment will have to rely on this source of 
knowledge.

%Another major player in the field that should be mentioned
%is Japan, which entertained for a while the prospect of building
%a linear $e^+e^-$ collider as a precision Higgs measurement 
%achine. However, at present the future prospect of this 
%proposal is in doubt. 

Other major players in the field of accelerator physics
that should be mentioned
are Japan and Russia.  Japan entertained for a while the prospect of building
a linear $e^+e^-$ collider as a precision Higgs measurement 
machine. However, at present the future prospect of this 
proposal is in doubt. Russia has diverse expertise in accelerator physics
in Dubna and Novosibirsk. Here again CERN provides the guidebook
for cooperation with diverse international entities and CERN provides
the guidelines for policy in delicate international situations. In this
respect, it can be envisioned that participation will be put on hold
of any country in active disagreeable dispute with any of its neighbours,
until such matters are resolved satisfactorily.

The current planning of new collider experiments prioritise the
development of new technology and the precision measurements of the
Higgs properties over reaching into new energy regimes from whence 
the discovery of new physics is likely to come. Exploring the 
new energy regimes is envisioned to come at a second stage in the 
2060s or 2070s.

However, there is an alternative route to follow, which is
particularly attractive to new players in the field. In the 
1990s the US planned a Supeconducting Super Collider (SSC) experiment
that was supposed to operate at 40TeV CoM energy, {\it i.e.} at 
nearly three time the LHC energies. The decision by the American Congress 
on the Texas site proposal 
of the experiment was taken in Autumn 1988. It was initially
supposed to become operational in 1996 which was later pushed to 1999.
It was nearly scrapped by the American congress in Summer 1992 but 
was miraculously saved. Part of the Opposition to the project was due to
the cost increase from an estimate of 6B to 11B USD. The main reason 
was a change of the magnet design from 4cm to 5cm aperture. In October 
1993 the American congress scraped the project after more than 20\%
of the 87.1km tunnel was dug and about 2B USD were spent. 

Our proposal is therefore of an Upgraded Superconduncting Super Collider (USSC). 
Namely to use the Original SSC (OSSC) design as a bench mark. The OSSC 
was planned on using 6.6T magnets with 5cm bore and 87.1km circular
ring. It was supposed to deliver proton beams with 20TeV per beam or 
40TeV CoM energy. As mentioned above, the decision on the Texas site 
was taken in October 1988 and it was supposed to start operation in 
the 1996--1999 period. It was cancelled in October 1993 due primarily to the cost increase from 6B to 11B USD. 

One can therefore use the LHC magnet technology 
to build an Upgraded SSC collider with similar specifications. Thirty years after the OSSC cancellation, with improved magnet
technology, 8-10T magnets with 5cm bore are feasible. 
With 90--100km tunnel, CoM in the 50--60TeV range can 
be foreseen. Given that the LHC magnets operate at 8.3T and 
can operate at 9--10T, the envisioned collider can
be characterised as FCC--LHC, {\it i.e.} an FCC experiment
that uses LHC magnet technology, rather than the new, 
yet unproven technology that uses Niobium--Tin alloys. 
Given the timeline of the OSSC from 1988 to 1999, the USSC
can be realised in 10--15 years from decision to completion, 
{\it i.e.} it can become operational by the late 2030s. 
From the Snowmass 2021 estimates, one can estimate 
the cost of the project to be of the order of 
10B--20B USD. The USSC will be able to do bread and butter
Standard Model physics, {\it i.e.} it will be able to 
measure the Standard Model parameters to better precision
than they are known today, but its main advantage is that 
it will be a discovery machine, {\it i.e.} it will probe the new
energy regimes beyond those that are currently being 
studied at the LHC. 

We then come to the inevitable question of where can
the USSC be built? At the time of writing all four 
main players, China, Europe, Japan and the US, 
have initiated well established processes to 
determine their future accelerator physics 
programs. While the option of an FCC experiment
using LHC magnet technology has been discussed,
it is not a front runner in the deliberations. 
The reasons are varied. The leading international 
laboratory, CERN, has a well established program in the
High Luminosity--LHC (HL--LHC) that will run until the 
mid to late 2030s. It is sensible to maximise the output 
from the beautifully running LHC. CERN then plans to 
follow its LEP--LHC playbook experience. Namely, first dig
a tunnel for a precision lepton collider, the FCC--$e^+e^-$
collider, and use the same tunnel for a proton--proton collider, 
the FCC--$hh$. CERN is a marvel to behold. A beacon in a dark
void. One cannot fault CERN with anything that it does, and 
whatever CERN decides to do -- it will be done. 

The other players are not at the same organisational 
maturity as CERN, but in principle could have picked the 
USSC. They have the basic technological infrastructure to 
develop it, albeit not with the same level of experience. 
The caveat is that while there is a good prospect 
to discover new physics at the USSC, it is not guaranteed. 
For that reason, in regions where the basic technological 
infrastructure already exists, considering the overall
USSC projected cost, it makes sense to prioritise 
the technology development, {\it i.e.} the development 
of new superconducting alloys that allow the production of
magnets with higher magnetic fields. Such magnets can
be used in many applications, such as medical applications; 
levitated faster moving trains; and energy storage; among 
others. 

There is therefore an opportunity for new players to come 
into the field. We therefore propose that the USSC
can be built as a Middle East project in the SESAME 
laboratory in Jordan. The SESAME laboratory provides 
the initial organisational infrastructure, which can be 
further developed along the lines of CERN. Regional 
countries, like Saudi--Arabia and the Arab Emirates 
have the financial muscle to carry out the project. 
The SESAME site, mimicking the location of CERN at 
a central European location, yet at a host country, 
which is not and has not been, one of those aspiring for a
domineering position over its neighbours. Thus, it serves as
a host for all regional countries and beyond, promoting  
cooperation and collaboration in an enterprise of common human 
interest. 
The potential benefit in terms of technological 
development and global prestige is enormous.
Additionally, the acquired expertise in super--conducting 
magnet technology and manufacturing has wide range of 
applications, from medical imaging through levitated trains,
to energy storage and more. 
If in 
Europe or the US, the USSC cost can be estimated between
10B--20B USD, its cost as a ME project will likely be
five times that, {\it i.e.} 50B--100B USD. CERN
experience and guidance will be vital for the success
of the project and to alleviate CERN concern about 
the competition, agreements can be reached with
the participating countries for a common benefit. 
The project will be developed under the leadership of the Arab 
Physical Society and its president Professor Shaaban Khalil.
Professor Khalil established a thriving group in 
high energy physics at the Zeweil City University in Cairo and 
is a member of the CMS collaboration of the LHC experiment at CERN. 
One may think of this proposal as FCC--LHC, {\it i.e.} as a
proton--proton Collider with 90--100km tunnel with LHC magnet 
technology. The current state of affairs in the field provides
an excellent opportunity for new players to come in and potentially
reap tremendous benefits both in terms of technological and 
industrial developments, as well as the potential of 
breakthrough discoveries and the prestige that they bring. 

The USSC will require substantial financial investment and 
industrial development to produce the components of the collider. 
The mere production of 90--100km of superconducting magnets
with the cryogenic cooling systems will necessitate the appropriation 
of substantial industrial capacity. But the mere technological 
development is not the main motivation to propose the USSC. 
The physics case is strong and offers the potential of 
breakthrough discoveries. Furthermore, given the state
of our understanding of the fundamental particles 
and their interactions, the USSC is the optimal 
machine to build at this point, and in particular it 
is the optimal machine to be built by new players 
coming into the field without earlier commitments. 

\begin{figure} 
\begin{center}
\includegraphics[width=7.cm]{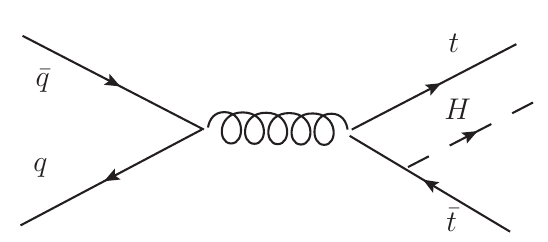}
\caption{$q{\bar q}$ annihilation into $t{\bar t}$--Higgs}
\label{qqtttbarhiggs}
\end{center}
\end{figure}

\begin{figure} 
\begin{center}
\includegraphics[width=7.cm]{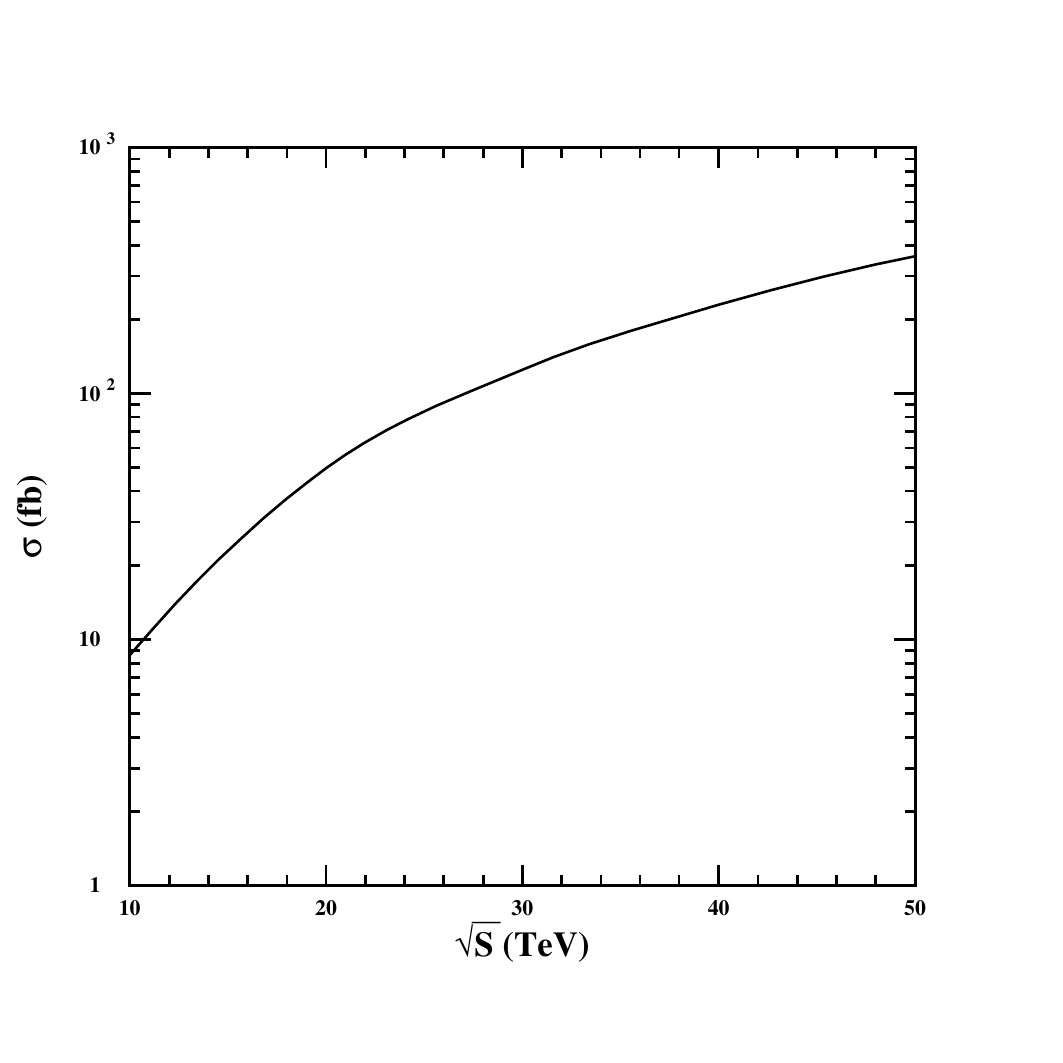}
\caption{Associated production of Higgs boson and $t{\bar t}$ 
at the LHC (figure taken from arXiv:1508.06416 \cite{1509.06416})}
\label{cssm}
\end{center}
\end{figure}

The physics case of the USSC experiment must first 
of all be based on the bread \& butter physics 
that it can do with the Standard Model parametrisation. 
The first task of the experiment will be to improve the
measurement of the Standard Model parameters and reduce 
the error bars. One example is the process of proton--proton
collisions into top--anti--top--Higgs in the final state
that provides a direct measurement of the $t{\bar t}h$ vertex
in the Standard Model Lagrangian. The process is shown 
in figure \ref{qqtttbarhiggs}.
Figure \ref{cssm} shows the cross section for associated production of 
the Higgs boson and top--anti--top quarks at $pp$ collider as a
function of the Centre of Mass (CoM) energy. Going from LHC 
energy of 14TeV to 50TeV increases the cross section by roughly an 
order of magnitude. This is an example of a process that will be 
accessible at a 50TeV $pp$ collider but might not be accessible 
in precision lepton colliders with energy reach between 250--500GeV. 
The associated production in figure \ref{qqtttbarhiggs} provides
a direct measurement of the crucial top quark Yukawa coupling which 
provides an important window to the physics of the Standard Model 
and beyond. The process highlighted in figures \ref{qqtttbarhiggs}
and \ref{cssm} is a mere solitary example, but there are of course 
many more, that illustrate the potential gain of going from LHC 
energy of 14TeV CoM energy to 50TeV envisioned at the USSC.  

The USSC will therefore do substantial amount of Bread \& Butter 
physics measurements and improve the measurement of the 
Standard Model parameters by an order of magnitude or so. 
Furthermore, the USSC will be able to measure and constrain 
parameters in the Higgs sector, like the cubic Higgs coupling 
that are not accessible at the LHC. However, the main selling 
point of the USSC is that it will probe the physics beyond 
the Standard Model and in particular the sector which is 
associated with electroweak symmetry breaking. 

\begin{figure} 
\begin{center}
\includegraphics[width=7.cm]{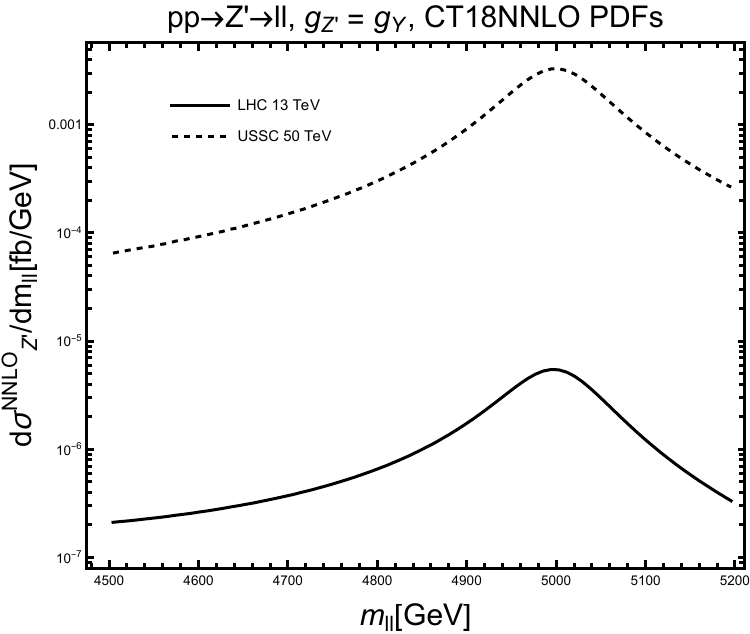}
\caption{$Z^\prime$ production and leptonic decay at 14TeV and 50TeV
(figure taken from arXiv:2309.15707 \cite{USSC}) }
\label{zpatpp}
\end{center}
\end{figure}

To illustrate the sensitivity to new physics Beyond the Standard Model
I discuss the production of $Z^\prime$ vector boson in a string
derived model \cite{USSC}. The gauge interactions in the Standard Model 
are mediated by elementary particles with particular properties, and 
we posit the existence of a new such interaction that has not yet
been observed experimentally. This entails that the spacetime 
vector boson that mediate the new interaction is heavier than
the energies reached in contemporary collider experiments
but may be produced at the energy scale accessible at the 
USSC. It is particularly interesting to explore the existence of 
such a new spacetime symmetry in the context of string theory. 
String theory smooths out the divergences that arise 
when one tries to combine the Standard Model of particle 
physics with gravity. It achieves that because in string 
theory elementary particles are not idealised point 
particles, but rather have an extended internal dimension. 
However, the natural scale where this internal structure will
be revealed is far removed from the electroweak scale.
String theory, however, gives rise to models that reproduce
the main features of the Standard Model and its supersymmetric
extension and we can connect between these string models 
and experimental data by interpolating their parameters to the 
contemporary energy scales. In that story, the existence of an 
additional spacetime vector boson beyond the Standard Model
can explain why the electroweak scale is so much lower than the 
natural string scale. We can then calculate the production of the 
extra vector boson at the LHC and in future $pp$ colliders. 
Figure \ref{zpatpp} illustrates the production of a $Z^\prime$
vector boson at the LHC and at the USSC. From the figure it is 
seen that three orders of magnitude in the cross section
are gained by going from LHC energy of 14TeV to the USSC
energy of 50TeV. At the USSC we will therefore gain an 
increase by a few orders of magnitude to the number
of events if new resonances, as is widely anticipated, 
exist in the energy regime beyond 14TeV. New discoveries
and the riches that they bring await those that are 
bold enough to venture into the unknown.

Building the USSC is an enormous technological challenge in any 
locality, let alone at the SESAME facility, where the technical 
know--how does not exist at the moment. It is an enormous, technological,
logistical and societal challenge that at the moment could 
only be achieved in few places. Chief among them is CERN.  
CERN guidance and expertise will be vital in any future 
accelerator particle physics experiment. If not directly managed, the 
USSC will benefit from a close relationship with CERN and its 
leadership. The development of the accelerator basic design and 
the physics case will have to engage the worldwide community in a
global effort. The UK, in general, and the Northwest of England, 
in particular, have a long and established tradition in 
accelerator physics and can, for example, in collaboration with the
Cockcroft and Adams institutes for accelerator physics, assist in the
training of future generations of accelerator physicists and engineers. 

As we look to the future of humanity's quest for understanding
the laws that govern the basic matter and interactions in the 
smallest and largest length scales in the subatomic and cosmological 
domains, and the instruments that will be needed on that quest, we 
are reminded of our forebears and the journeys that they have taken
on that quest. Columbus in his time sailed west seeking a route to 
India. Unwittingly, he discovered America and changed the course of
history. Similarly, the
USSC will seek to measure the parameters of the Standard Model
to better precision and to do Bread \& Butter Standard Model physics. 
It may, however, discover new physics Beyond the Standard Model
that is associated with the Electro--Weak Symmetry Breaking 
Mechanism, which is widely expected to exist. Given the 
OSSC timeline, it may do so by 2040. 

\bibliography{paperbib}
\bibliographystyle{JHEP}

\end{document}